\documentstyle[12pt]{article}
\begin{document}
\noindent
\begin{center}
{\LARGE {\bf  SCALAR TENSOR THEORIES AND\\ HADAMARD STATE CONDITION\\}}
\vspace{2cm}
${\bf H.~Salehi}$\footnote{e-mail:h-salehi@cc.sbu.ac.ir.},
${\bf Y.~Bisabr}$\footnote{e-mail:y-bisabr@cc.sbu.ac.ir.},
${\bf H.~Ghafarnejad}$ \\
\vspace{0.5cm}
{\small {Department of Physics, Shahid Beheshti University, Evin,
Tehran 19839,  Iran.}}\\
\end{center}

\begin{abstract}
The Hadamard state condition is used to analyze
the local constraints on the two-point function of a quantum  field
conformally coupled to a background geometry.
Using these constraints we develop a
scalar tensor theory which controls the coupling
of the stress-tensor induced by the two-point function  of the quantum
field to the conformal class
of the background metric. It is then argued that
the determination of the state-dependent part of the two-point
function is connected with the determination of a
conformal frame. We comment on
a particular way to relate the theory to a specific conformal
frame (different from the
background frame) in which the large scale
properties are brought into focus. \vspace{30mm}
\end{abstract}
%%%%%%%%%%%%%%%%%%%%%%%%%%%%%%%%%%%%%%%%%%%%%%%%%%%%%%%%%%%%%%%%%%%%%%%%%%%%%%%

\section{Introduction}
The essential feature of scalar tensor theories, such as Brans-Dicke theory,
is to generalize general relativity and bring it into accord
with Mach's principle (the origin of physical properties of space is in the matter
contained therein \cite{1} ). These theories are not completely geometrical since
the gravitational effects are described by a scalar field as well as a metric
tensor. In fact, the global distribution of matter affects the local gravitational
properties through the emergence of a scalar field.
The implementation of this interrelation between global and
local properties
of matter in quantum field theory, as
demanded by Mach's principle, is a complicated problem. Some
ideas in this direction can be found in \cite{2,3}.\\
In a simplified picture one can expect that the role of a
scalar tensor theory
may be of importance for improving our knowledge on
the local properties of a linear quantum field propagating
in a gravitational background, in particular the local properties of the
quantum stress-tensor induced by the two-point function of the quantum field.
The present paper deals with the consideration of this issue.
In specific terms, we study a model in which the
local properties of
a linear quantum field conformally coupled
to a gravitational background
is affected both by the local geometry and
a conformal invariant scalar field derived from the
state (boundary)-dependent part of the two-point function. To arrive at this
model we basically take into account
the local constraints imposed on the two-point function by
the Hadamard state condition.
In this context there is a problem concerning the specification
of the state-dependent part of the two-point function.
In our presentation we establish a connection between this problem
and the problem of the determination of a conformal frame .\\
To avoid any confusion at the outset,
we should note that the scalar tensor theory we wish to consider
is meant only to provide
an analytical
mean to determine the general properties of a quantum stress-tensor that can
consistently be coupled to conformally related background metrics,
and in this respect its interpretation differs from
the standard interpretation of such theories as alternative theories of
gravitation.\\
The organization of this paper is as follows: In section 2 we present the
Hadamard prescription and review the derivation of the local constraints
on the state-dependent part of two-point function of a linear scalar quantum
field conformally coupled to gravity.
In section 3, we present a way to use a conformally invariant
scalar field for analyzing
the state-dependent part of the
two-point function. It is shown that the
implications of the resulting scalar tensor theory for the stress-tensor
are in accord with
the standard predictions of the renormalization theory.
In section 4, we make some general remarks on the existence of
an alternative frame in which the trace of the stress-tensor
is determined by a cosmological constant rather than the usual anomalous
trace. The existence of this frame indicates that
the state-dependent part of the two-point functions may have some
large scale characteristics which are basically not present
in the conformal frame determined by the local characteristics.
Similar arguments were discussed previously in a
different context \cite{3}.
%%%%%%%%%%%%%%%%%%%%%%%%%%%%%%%%%%%%%%%%%%%%%%%%%%%%%%%%%%%%%%%%%%%%%%%%%%%%%%%

\section{Hadamard state condition}
We consider a free scalar quantum field $\phi(x)$ propagating in a
curved background spacetime with the action functional \cite{4}
(We use the conventions of Hawking and Ellis \cite{5} for the
signature and the sign of curvature)
\begin{equation}
S[\phi]=-\frac{1}{2} \int d^{4}x g^{1/2}(g_{\alpha\beta} \nabla^{\alpha}\phi
\nabla^{\beta}\phi+\xi R \phi^{2}+m^{2} \phi^{2})
\end{equation}
where $m$ and $\xi$ are parameters, and $R$ is the scalar curvature
(In the following the semicolon and $\nabla$ indicate covariant differentiation).
This gives rise to the field equation
\begin{equation}
(\Box-m^{2}-\xi R)\phi(x)=0
\end{equation}
The choice of the parameters $m$ and $\xi$ depends on the particular type
of coupling. For example, the minimal coupling corresponds to $m$=0, $\xi=0$
and the conformal coupling (in four dimensions) corresponds to $m$=0,
$\xi=1/6$ .
A state of $\phi(x)$ is characterized by a hierarchy of
Wightman-functions (n-point functions)
\begin{equation}
\langle \phi(x_1),...,\phi(x_n) \rangle
\end{equation}
We are primarily interested in those states which reflect the
intuitive notion of a "vacuum". For this aim, we may restrict ourselves
basically to
quasi-free states, i.e. states for which the truncated
n-point functions vanish for $n>2$ (In a linear theory this property
is shared by the vacuum state
of Minkowski space).
Such states may be characterized
by their two-point functions.
In a linear theory the antisymmetric part of the two-point function
is common to all states in the same representation. It is just
the universal commutator function. Thus, in our case all the relevant
informations about the state-dependent part of the two-point function are
encoded in its symmetric part, denoted in the following by $G^{+}(x,x')$,
which satisfies Eq.(2) in each argument. Equivalence
principle suggests that the
leading
singularity of $G^{+}(x,x')$ should have a close correspondence to the singularity
structure of the two-point function of a free massless field in Minkowski
space. In general the entire singularity of $G^{+}(x,x')$ may have a more
complicated structure. Usually one assumes that $G^{+}(x,x')$ has a singular
structure represented by the Hadamard expansions. This means that in a normal
neighborhood of a point $x$ the function $G^{+}(x,x')$ can be written \cite{6,7,8}
as
\begin{equation}
G^{+}(x,x')=\frac{1}{8 \pi^{2}} \{\frac{\Delta^{1/2}(x,x')}{\sigma(x,x')}+
V(x,x')~\ln\sigma(x,x')+W(x,x')\}
\end{equation}
where $2\sigma(x,x')$ is the square of the distance along the geodesic
joining $x$ and $x'$ and $\Delta(x,x')$ is the Van Vleck determinant
\begin{equation}
\begin{array}{ll}
\Delta(x,x')=-g^{-1/2}(x) Det\{-\sigma_{;\mu\nu _{'} }\}g^{-1/2}(x')\\
g(x)=Det~ g_{\alpha\beta}
\end{array}
\end{equation}
The functions $V(x,x')$ and $W(x,x')$ are regular and have the
following representations as power series
\begin{equation}
V(x,x')=\sum_{n=0}^{+\infty} V_{n}(x,x') \sigma^n
\end{equation}
\begin{equation}
W(x,x')=\sum_{n=0}^{+\infty} W_{n}(x,x') \sigma^n
\end{equation}
in which the coefficients are determined by applying Eq.(2) to
$G^+(x,x')$, yielding the recursion relations
\begin{equation}
(n+1)(n+2)V_{n+1}+(n+1)V_{n+1;\alpha} \sigma^{;\alpha}-(n+1)V_{n+1} \Delta^{-1/2}
\Delta^{1/2}_{;\alpha} \sigma^{;\alpha}+\frac{1}{2}(\Box-m^2-\xi R)V_n =0
\end{equation}
$$
(n+1)(n+2)W_{n+1}+(n+1)W_{n+1;\alpha}\sigma^{;\alpha}-(n+1)W_{n+1}\Delta^{-1/2}
\Delta^{1/2}_{;\alpha} \sigma^{;\alpha}
+\frac{1}{2}(\Box-m^2- \xi R)W_{n}
$$
\begin{equation}
~~+(2n+3)V_{n+1}+V_{n+1;\alpha}\sigma^{;\alpha}
-V_{n+1} \Delta^{-1/2}\Delta^{1/2}_{;\alpha}\sigma^{;\alpha}=0
\end{equation}
together with the boundary condition
\begin{equation}
V_{0}+V_{0;\alpha}\sigma^{;\alpha}-V_{0}\Delta^{-1/2}\Delta^{1/2}_{;\alpha}
\sigma^{;\alpha}+\frac{1}{2}(\Box-m^2- \xi R) \Delta^{1/2}=0
\end{equation}
From these relations one can determine the function $V(x,x')$ uniquely
in terms of local geometry. Therefore it takes the same universal form for
all states. But the biscalar $W_{0}(x,x')$ remains arbitrary. Its specification
depends significantly on the choice of a state and may be regarded as the imposition of a boundary
condition. However there is a general constraint on $W_{0}(x,x')$
which can be obtained from the symmetry condition of $G^{+}(x,x')$ together
with the following dynamical equation which can be obtained using (2),(4) and
(6) \cite{9,10}
\begin{equation}
( \Box -m^2- \xi R)W(x,x')=-6v_{1}(x)+2v_{1;\alpha} \sigma^{;\alpha}+
0(\sigma)
\end{equation}
where
\begin{equation}
v_{1}(x)=\lim_{x'\rightarrow x}V_{1}(x,x')=\frac{1}{720}\{ \Box R-R_{\alpha\beta}
R^{\alpha\beta}+R_{\alpha\beta\delta\gamma}R^{\alpha\beta\delta\gamma}\}
\end{equation}
To get this constraint we first expand the symmetric function $W(x,x')$
into a covariant power series, namely \cite{9,10,11}
\begin{equation}
W(x,x')=W(x)-\frac{1}{2}W_{;\alpha}(x) \sigma^{;\alpha}+\frac{1}{2}W_{\alpha \beta}(x)
\sigma^{;\alpha} \sigma^{;\beta}+\frac{1}{4}\{\frac{1}{6}W_{;\alpha\beta\gamma}(x)-
W_{\alpha\beta;\gamma}(x)\}\sigma^{;\alpha}\sigma^{;\beta}\sigma^{;\gamma}+
0(\sigma^{2})
\end{equation}
We may insert this into Eq.(11) and compare term
by term up to the third order in $\sigma^{;\alpha}$ to obtain
\begin{equation}
W^{\gamma}_{\gamma}(x)=( \xi R+m^2)W(x)-6v_{1}(x)
\end{equation}
\begin{equation}
[W_{\alpha\beta}(x)-\frac{1}{2}g_{\alpha\beta}W^{\gamma}_{\gamma}(x)]^{;\alpha}
=\frac{1}{4}(\Box W(x))_{;\beta}-\frac{1}{2}m^2W_{;\beta}(x)+2v_{1}(x)_{;\beta}
+\frac{1}{2}R_{\alpha\beta}W^{;\alpha}(x)-\frac{1}{2}\xi RW_{;\beta}(x)
\end{equation}
Then using the covariant expansion of the symmetric function $W_{0}(x,x')$
\begin{equation}
W_{0}(x,x')=W_{0}(x)-\frac{1}{2}W_{0;\alpha}(x) \sigma^{;\alpha}+\frac{1}{2}W_{0\alpha \beta}(x)
\sigma^{;\alpha} \sigma^{;\beta}+0(\sigma^{3/2})
\end{equation}
together with Eqs.(7),(9) and (13), we get \cite{9,10}
\begin{equation}
W(x)=W_{0}(x)
\end{equation}
\begin{equation}
W_{\alpha\beta}(x)=(W_{0\alpha\beta}(x)-\frac{1}{4}g_{\alpha\beta}
W^{\gamma}_{0\gamma}(x))+\frac{1}{4}[(m^2+\xi R)W_{0}(x)-6v_{1}(x)]
g_{\alpha\beta}
\end{equation}
Substituting (17) and (18) into (15) leads to
$$
[W_{0\alpha\beta}(x)-\frac{1}{4}g_{\alpha\beta}W^{\gamma}_{0\gamma}(x)]^{;\alpha}
=\frac{1}{2}v_{1;\beta}(x)+\frac{1}{4}(\Box W_{0}(x))_{;\beta}-
\frac{1}{4}m^2W_{0;\beta}(x)+\frac{1}{2}R_{\alpha\beta}W_{0}^{;\alpha}(x)
$$
\begin{equation}
~~~+\frac{1}{4}\xi [R_{;\beta}W_{0}(x)-RW_{0;\beta}(x)]
\end{equation}
This equation is a general constraint imposed on the state-dependent
part of the two-point function. The function $W_{0}(x)$
may be considered as
arbitrary, but once a specific assumption has been made on the
form of $W_{0}(x)$,
the equation (19) acts as a constraint on
$W_{0\alpha\beta}(x)$.\\
We should note that the constraint (19) is, in principle, the first member
of a hierarchy of constraints, because
we have used the covariant expansion $W_{0}(x,x')$ only
up to the second order in $\sigma^{;\alpha}$. Thus,
in general, there are some additional
constraints on the higher order expansion terms.
In our analysis we shall neglect these higher
order constraints.
Such a limitation is suggested by dimensional arguments
because the second order expansion terms of $W_{0}(x,x')$ has already the
physical dimension of a stress-tensor.
%%%%%%%%%%%%%%%%%%%%%%%%%%%%%%%%%%%%%%%%%%%%%%%%%%%%%%%%%%%%%%%%%%%%%%%%%%%%%%%

\section{The conformally invariant scalar field}

In the case of conformal coupling a local Hilbert space
would in general exhibit an essential
sensitivity to the pre-existing
local causal structure of space-time which in the present
case is determined by the conformal class of the background metric.
By implication,
this causal structure should act as the basic input for the characterization
of the local states.
Since the conformal
transformations leave the causal structure unchanged
we expect, in particular, that
an essential ambiguity, related to conformal transformations, should enter
the dynamical specification of the state-dependent part of the
two-point function.
Thus, in the case of the conformal coupling, it is suggestive to
develop
a dynamical model in which the conformal symmetry
acts as a fundamental symmetry in the specification
of the two-point function, in particular the function $W_{0}(x)$.
In the following we shall use the constraint (19)
to develop a dynamical model along this line.
We first start
with the explicit form of the constraint (19) in the case of
conformal coupling, namely
$$
[W_{0\alpha\beta}(x)-\frac{1}{4}g_{\alpha\beta}W_{0\gamma}^{\gamma}(x)
-\frac{1}{2}g_{\alpha\beta}v_{1}(x)-\frac{1}{4}g_{\alpha\beta}
\Box W_{0}(x)]^{;\alpha}~~~~~~~~~~~~~~~~~~~~~~~~~~~~~~~~~~~~~~~~~~~~~~~~~~~~~~
~~~~~~~~~~~~~~~~~~~~~~~~~~~~~~~~~~
$$
\begin{equation}
~~~~~~~~~~~~~~~~~~~~~~~~~~=\frac{1}{2}R_{\alpha\beta}W_{0}^{;\alpha }(x)+\frac{1}{24}(R_{;\beta}
W_{0}(x)-RW_{0;\beta}(x)).
\end{equation}
One can use the Bianchi identity
\begin{equation}
R_{\alpha\beta}^{;\alpha}=\frac{1}{2}R_{;\beta}
\end{equation}
and the differential identity
\begin{equation}
\Box(W_{0;\beta}(x))=(\Box W_{0}(x))_{;\beta}+R_{\alpha\beta}W_{0}^{;\alpha}(x)
\end{equation}
to show that (20) can be written as a total divergence
\begin{equation}
\Sigma_{\alpha\beta}^{;\alpha}=0
\end{equation}
where
$$
\Sigma_{\alpha\beta}=(W_{0\alpha\beta}(x)-\frac{1}{4}g_{\alpha\beta}
W_{0\gamma}^{\gamma}(x))-\frac{1}{6}(R_{\alpha\beta}-\frac{1}{4}Rg_{\alpha\beta})W_{0}(x)
-\frac{1}{3}(W_{0;\beta\alpha}(x)-\frac{1}{4}g_{\alpha\beta}
\Box W_{0}(x))
$$
\begin{equation}
-\frac{1}{2}g_{\alpha\beta}v_{1}(x) ~~~~~~~~~~~~~~~~~~~~~~~~~~~~~~~~~~~~~~~~~~~~~~~~~~~~~~~~~~~~~~~~~~~~~~~~~~~
\end{equation}
Now, the basic input is to subject in (23) the choice of
$W_{0}(x)$ to the condition
\begin{equation}
W_{0}(x)=\psi^{2}(x)
\end{equation}
where $\psi(x)$ is taken to be a conformally invariant
scalar field
coupled to the gravitational background, so that its dynamical equation is
\begin{equation}
(\Box-\frac{1}{6}R)\psi=0.
\end{equation}
For a given Hadamard state the field $\psi$ may be interpreted as measuring
the one-point function of the quantum field $\phi$.
The conformal invariance of $\psi$ ensures that there exists
no pre-assigned dynamical configuration for the one-point function in a local
Hilbert space. This is indeed a desirable characteristic of a linear theory.\\
Technically, the merit of introducing the field $\psi$ is that the
tensor $\Sigma_{\alpha\beta}+\frac{1}{2}g_{\alpha\beta}v_{1}(x)$,
which is traceless due to (24), may now be related to the
conformal stress-tensor of $\psi$, namely
\begin{equation}
\Sigma_{\alpha\beta}+\frac{1}{2}g_{\alpha\beta}v_{1}(x)= T_{\alpha\beta}[\psi]
\end{equation}
where the conformal stress-tensor
$T_{\alpha\beta}[\psi]$ is given by \cite{12}
\begin{equation}
T_{\alpha\beta}[\psi]=(\frac{2}{3}\nabla_{\alpha}\psi\nabla_{\beta}\psi-\frac{1}{6}
g_{\alpha\beta}\nabla_{\gamma}\psi\nabla^{\gamma}\psi)-\frac{1}{3}
(\psi\nabla_{\alpha}\nabla_{\beta}\psi-g_{\alpha\beta} \psi\Box \psi)
+\frac{1}{6}\psi^2 G_{\alpha\beta}
\end{equation}
in which $G_{\alpha\beta}$ is the Einstein tensor. The tensor $T_{\alpha\beta}$
is traceless due to the dynamical equation (26).
The meaning of the relation (27) is that it defines a formal
prescription which allows us to relate
the tensor $W_{0\alpha\beta}(x)$ in (24)
to the function $W_{0}(x)$ and the metric tensor
$g_{\alpha\beta}$, so it characterizes a criterion to select the class of
admissible Hadamard states.
Taking into account (28) we can write this criterion as
\begin{equation}
G_{\alpha\beta}-3\psi^{-2}g_{\alpha\beta}v_{1}(x)=
6\psi^{-2}(\Sigma_{\alpha\beta}+\tau_{\alpha\beta}(\psi)).
\end{equation}
Here $\tau_{\alpha\beta}(\psi)$,
is equal to $T_{\alpha\beta}[\psi]$
without the $G_{\alpha\beta}$-term, so it coincides up to a sign
with the so called modified energy-momentum (stress-) tensor \cite{13}.
Now, the basic strategy is to consider the tensor
$\Sigma_{\alpha \beta}$
as the quantum stress-tensor induced by the two-point function.
Our criterion can then be interpreted as a rule for relating the latter tensor
to the local background geometry, as reflected in (29).
The essential point is that
this rule is expressed in the form of a scalar
tensor theory in which the dynamics of
the scalar field $\psi$ makes substantially no distinction
between different frames in the conformal class of
the background metric.
The implication is that at the dynamical level all conformal frames may
be considered as equivalent.   \\
This conformal invariance reflects a basic connection
between the state-dependent part of the two-point
function and the pre-existing causal structure determined
by the background metric. In particular, it establishes a basic
connection
between the properties of a given physical state in a local Hilbert space
and those of a corresponding conformal frame. To see this in explicit
terms let us
consider a conformal
transformation
$$
\bar{g}_{\alpha\beta} =\Omega^{2}(x) g_{\alpha\beta}
$$
\begin{equation}
\bar{\psi}(x) = \Omega^{-1}(x) \psi(x)
\end{equation}
Due to (25), $W_{0}(x)$ would then transform as
\begin{equation}
\bar{W}_{0}(x) = \Omega^{-2}(x) W_{0}(x)
\end{equation}
It is now clear from (31) that
a given conformal frame may be characterized by the particular configuration
of $W_{0}(x)$ (or alternatively $\psi$) in that frame.
Therefore the problem of specification of $W_{0}(x)$ for a
given physical state is basically connected with the problem of
determination of a conformal frame. In particular,
different states characterized by conformally related
configurations of $W_{0}(x)$ should principally be supported
on different conformally related metrics.
The same conclusion
holds for their stress-tensors.\\
At this point we make a general remark concerning the consistency
of our results with the standard prediction of the renormalization theory.
Focusing ourselves to the two-point function on the background
metric we can take the trace of
(27), to obtain
\begin{equation}
\Sigma_{\alpha}^{\alpha}=-2v_{1}(x)
\end {equation}
This together with (23) characterize the general properties of
the quantum stress-tensor on the background metric.
These properties
are consistent with the well-known
results of the renormalization theory \cite{14}
and $v_{1}(x)$ is actually the function that determines what is
commonly known as the trace anomaly. In our presentation
this quantum anomaly requires a somewhat
distinct behavior of the scalar field $\psi$.
In fact, according to (23) and (27) and due to the nonvanishing
trace anomaly,
the tensor $T_{\alpha\beta}[\psi]$, which may be considered as
the stress-tensor of the field
$\psi$,
appears not to be conserved on the background metric, requiring
the dynamical properties of $\psi$ on the background metric not to fit
in with the properties
of a diffeomorphism invariant action characterizing a
C-number (classical) field.
But it is necessary to stress
that this behavior does not appear to be a physical contradiction
in the present case.
Actually, the scalar field $\psi$ which characterizes the local property
of the two-point function may in general change
its configuration if one varies the two-point function within
a local Hilbert space.
Therefore, in general it may not act as a C-number field within a local
Hilbert space. By implication,
the standard results of a diffeomorphism invariant action may not be
applied to $\psi$. We note that a similar process
of assigning non-diffeomorphism invariant properties to
a local Hilbert space has been previously discussed
in the context of generally covariant quantum field theory [2].

%%%%%%%%%%%%%%%%%%%%%%%%%%%%%%%%%%%%%%%%%%%%%%%%%%%%%%%%%%%%%%%%%%%%%%%%%%%%%%%

\section{ $\Lambda$-frame}

The conformal symmetry which was established in the local specification
of $W_{0}(x)$ would imply
that locally the stress-tensor $\Sigma_{\alpha\beta}$ can be related
to different conformal frames.
Thus the question arises as to which frame
should be considered as a physical frame.
To deal with this question it is necessary to emphasize the role
of the superselection rules which characterize
the boundary conditions imposed on the physically realizable states
and the corresponding Hilbert spaces.
In general, the identification of a conformal frame as a physical
frame depends on the particular superselection rule one wishes to apply.
Of direct physical significance, in the present case, is a
superselection rule that tells us how
a local Hilbert space is linked to the large scale
boundary conditions imposed on physical states.
If the latter conditions correspond to the presence of large
scale distribution of matter
whose energy density is measured by a cosmological constant,
one may
subject the determination of a conformal frame (alternatively a local
Hilbert space) to the asymptotic correspondence between the anomalous trace
and a nonvanishing
cosmological constant at sufficiently large
spacelike distances.
In general, this condition may not be realized in the underlying background
frame, so in this case the physical frame is expected to be different from
the background frame.\\
This observation opens a way to study the transition from the
local characteristics of physical states in a local Hilbert space to
the large scale characteristics, which is expected to be of particular
importance for
establishing the large scale gravitational coupling of
physical states in a local Hilbert space.
Since by such a transition
the small distance properties are no more important, we may take
the overall correspondence between
the anomalous trace and
a nonvanishing cosmological constant everywhere
as the defining characteristic of
a local conformal frame which, by implication,
acts as the physical frame if one focuses on
large scale characteristics of physical states in the presence
of large scale distribution of matter.
For the construction of this frame one needs only to
apply a conformal transformation
to the background frame which establishes
the correspondence between the trace
anomaly and a nonvanishing cosmological constant.
Denoting the cosmological constant by $\Lambda$,
the corresponding
conformal factor may be taken to satisfy the equation
\begin{equation}
-3\Omega^{2}(x)\psi^{-2}v_{1}(\Omega^{2}(x)g_{\alpha\beta})=\Lambda
\end{equation}
Under this conformal transformation the equation (27) transforms to
\begin{equation}
\bar{\Sigma}_{\alpha\beta}-\frac{1}{6}\Lambda\bar{g}_{\alpha\beta}\bar{\psi}^2
=T_{\alpha\beta}[\bar{\psi}]
\end{equation}
or, equivalently
\begin{equation}
G_{\alpha\beta}(\bar{g}_{\alpha\beta})+\Lambda \bar{g}_{\alpha\beta}=
6\bar{\psi}^{-2}(\bar{\Sigma}_{\alpha\beta}+\tau_{\alpha\beta}(\bar{\psi})).
\end{equation}
Therefore in the new frame, which we call the $\Lambda$-frame, a
scalar tensor theory with a cosmological constant is obtained together with
Eq.(33) which is a complicated constraint on the conformal factor.
In the $\Lambda$-frame, contrary to the background frame, the
stress-tensor $\bar{\Sigma}_{\alpha\beta}$ may not be conserved.
However Eq.(34) implies that one can establish a conserved
stress-tensor by replacing $\bar{\psi}$ by a constant average
value $\bar{\psi}$=const.
In this case the usual features of general relativity can be established
in the $\Lambda$-frame. In particular, the tensor
$T_{\alpha\beta}[\bar{\psi}]$, which was found to be non-conserved in
the back-ground frame, becomes a multiple of the Einstein tensor,
so a conserved tensor.\\
For further investigation of the constraint (33), we write its explicit form on
the background metric. Using the conformal transformation of the function
$v_{1}(x)$ \cite{10} we find
$$
-e^{2\omega}\psi^{-2} \{3v_{1}(g_{\alpha \beta})+\frac{1}{240}[2R\Box\omega
+2R_{;\alpha }\omega ^{;\alpha }+6\Box(\Box\omega) +8[(\Box\omega )^2
$$
\begin{equation}
-\omega _{;\alpha \beta}\omega ^{;\alpha \beta}
-R_{\alpha \beta}\omega ^{;\alpha }
\omega ^{;\beta}-\omega ^{;\gamma }\omega _{;\gamma }\Box\omega
-2\omega _{;\alpha \beta}\omega ^{;\alpha }\omega ^{;\beta}]]\}=\Lambda
\end{equation}
where $\omega =-\ln\Omega$.
As an illustration we shall now apply (36) to study an
asymptotic relation between the
$\Lambda$-frame and a specific
background metric which we take to be
described by a Schwarzschild black hole.
In this case the function $v_{1}(x)$, which determines the trace
anomaly, reduces to
\begin{equation}
v_{1}(g_{\alpha \beta})=\frac{1}{720} R_{\alpha \beta\delta \gamma }
R^{\alpha \beta\delta \gamma }=\frac{1}{15} \frac{M^2}{r^6}
\end{equation}
where $M$ is the mass of the black hole.
Since the trace anomaly vanishes for $r\rightarrow \infty$,
one may generally expect that
for a sufficiently small $\Lambda$
there should be no distinction between the background and the $\Lambda$-frame
in a region far from the black hole event horizon.
That this behavior is dynamically allowed follows from the equation (36) as
we briefly demonstrate:
Let us restrict ourselves to the static case and assume that $\omega$ is only a function of $r$.
For $r>>2M$ the equation (36) takes then the form
\begin{equation}
\omega ^{''''}-4\omega ^{'2}\omega {''}=-40\Lambda\psi^{2}e^{-2\omega }
-8\frac{M^2}{r^6}~,~~~~~~~~~~~~~~~~~~r>>2M
\end{equation}
where prime indicates differentiation with respect to $r$. This equation
reveals that $\omega$ as a slowly varying function would be a solution for a
large value of $r$ and a sufficiently small cosmological constant.
In particular, for large values of $r$ an almost constant
conformal factor (close to one) can be used to
establish the correspondence
between the background frame and  the $\Lambda$-frame.

%%%%%%%%%%%%%%%%%%%%%%%%%%%%%%%%%%%%%%%%%%%%%%%%%%%%%%%%%%%%%%%%%%%%%%%%%%%%%%

\section{Summary and outlook}
For a quantum field conformally coupled to a gravitational background
we have presented a model in which the role of
a scalar tensor theory is emphasized
for studying the local constraints
imposed on physical states by the
Hadamard state condition. The corresponding scalar field is conformally
invariant and controls the coupling of the stress-tensor to
the conformal class of the background metric.
The predictions of this theory are
in accord with the standard results of the stress-tensor renormalization
if one chooses a conformal frame corresponding
to the background metric. We have emphasized that the choice of a
specific conformal frame as a physical frame must, in general, be subjected
to the superselection rules regulating the
coupling of a local Hilbert space to the physical conditions at
distant regions.
In this context we have discussed the possibility to consider
the theory in a distinguished frame, namely the $\Lambda$-frame, which
may act as the physical frame for the establishing
the large scale gravitational coupling of physical states in the presence
of large scale distribution of matter.
It is suggestive to link this
large scale gravitational coupling of physical states, reflected in the
$\Lambda$-frame, with their cut-off property
in the short distance scaling. A dynamical
cut-off theory of this type, if properly formulated,
would reflect one of the characteristic implication
of Mach's principle in quantum field theory.

\end{document}